\documentclass[twocolumn,showpacs,preprintnumbers,amsmath,amssymb,a4paper]{revtex4}

\usepackage{amsmath,amssymb,bbm,epsfig}

\newcommand{\be}{\begin{equation}}
\newcommand{\ee}{\end{equation}}
\newcommand{\ba}{\begin{array}}
\newcommand{\ea}{\end{array}}
\newcommand{\bqa}{\begin{eqnarray}}
\newcommand{\eqa}{\end{eqnarray}}

\newcommand{\n}{\nonumber}

\newcommand{\tr}{\mbox{Tr}}
\newcommand{\bra}[1]{\ensuremath{\langle #1 |}}
\newcommand{\ket}[1]{\ensuremath{| #1 \rangle}}
\newcommand{\prj}[1]{\ensuremath{| #1 \rangle \langle #1 |}}

\begin{document}

\title{An observable entanglement measure for mixed quantum states}

\author{Florian Mintert$^1$ and Andreas Buchleitner$^2$}

\affiliation{
$^1$Department of Physics, Harvard University,
17 Oxford Street, Cambridge Massachusetts, USA}
\affiliation{
$^2$Max-Planck-Institut f\"ur Physik komplexer Systeme,
N\"othnitzerstr. 38, 01187 Dresden, Germany}

\date{\today}

\begin{abstract}
We quantify an unknown mixed quantum state's entanglement 
by suitable, local parity measurements on its two-fold copy. The
associated observable qualifies as a generalized entanglement witness. 
\end{abstract}


\maketitle

Quantum entanglement is arguably the most bizarre and anti-intuitive feature
of quantum mechanics, and, furthermore, the key
ingredient for an upcoming quantum information technology. 
It is the cause of quantum-nonlocality, leading to `spooky action on
a distance' and the violation of Bell's inequalities, and opens
novel means of data encryption and communication, as well as the
efficient factorization of large numbers as a crucial prerequisite
of breaking cryptographic codes. 

Yet, despite its primordial importance, entanglement is hard to grasp:
There is so far no observable which allows for the direct measurement
inscribed into a given, arbitrary quantum state. Only indirect ways to
assess a given quantum state's degree of entanglement are available:
Either through {\em state-selective} entanglement witnesses, which are auxiliary observables
to identify predefined classes of entangled states -- i.e., some a
priori knowledge on the state to be detected is required, and other
classes of states with exactly the same entanglement properties may remain
unidentified. Or through quantum tomography -- the
experimental reconstruction of the full density operator $\varrho$ from the
measurement of a complete set of observables, followed by the
evaluation of some entanglement measure, which is in general a
nonlinear function of $\varrho$. While witnesses, though
efficiently implementable, are no reliable tool for all purposes, 
tomography implies a rapidly growing experimental
overhead as the Hilbert space dimension of the composite system under
study increases -- either through increasing subdimensions of the
subsystems, or through an increasing number of these. This rapidly
saturates experimental resources. 

Therefore, alternative strategies \cite{phorodecki03,bovino:240407} are urgently needed, since
experiments now succeed to control increasingly large quantum systems,
though meet a hard barrier when it comes to measure efficiently and in
real time the available amount of
entanglement as their central resource. In the present contribution,
we describe how few 
experimental measurements on a {\em two-fold copy} $\varrho\otimes\varrho$ of the
mixed state to be analysed provide a tight estimate of the
entanglement inscribed in $\varrho$, for bipartite systems
of arbitrary finite dimension. This defines a new strategy to
overcome the above impediments, and also yields a generalized
entanglement witness.

We start out with a short reminder of pure state entanglement and
the efficient measurement thereof, what will already fix the algebraic
structure which we will use in our subsequent generalization for mixed
states:
The {\em concurrence} $c(\Psi)$ of a finite dimensional bipartite
pure state 
$\ket{\Psi}$
can be expressed through the expectation value of the self-adjoint operator
$A=4P_-\otimes P_-$,
with respect to a {\em two-fold copy}
$\ket{\Psi}\otimes\ket{\Psi}$ of
$\ket{\Psi}$ \cite{pereswootters,brun04,mpc}:
\be
c(\Psi)=\sqrt{\bra{\Psi}\otimes\bra{\Psi}\ A\ \ket{\Psi}\otimes\ket{\Psi}}\ .
\label{conc_neu}
\ee
$P_-$ is the projector on the {\em antisymmetric} subspace of
the two copies of either subsystem. 
$c(\Psi)$ is directly accessible in laboratory experiments, through a
projective measurement of the antisymmetric component of
$\ket{\Psi}\otimes\ket{\Psi}$ in either subsystem, as 
recently demonstrated for twin photons \cite{expmc}. The
same recipe applies 
for higher-dimensional bipartite systems,
and for multipartite generalizations of concurrence \cite{mpc}, since
the algebraic structure of (\ref{conc_neu}) prevails.

We now want to generalize this measurement prescription for {\em mixed
states} of bipartite quantum systems. The key difficulty here stems
from the abstract definition of mixed state concurrence through the
``convex roof'' construction, which takes account of the
non-uniqueness of the pure state decomposition of an arbitrary density
matrix: Consequently, one has to determine the minimum average concurrence
$c(\varrho)=\inf\sum_ic(\phi_i)$
of all ensembles $\{\ket{\phi_i}\}$ that
describe the density matrix $\varrho$ \cite{eof}. This
distinguishes nonclassical from classical correlations typical of a
statistical mixture. The 
optimization problem renders the general evaluation of the convex roof
a hard mathematical task, and any direct approximation thereof will
yield an upper rather than a lower bound of the entanglement of
$\varrho$. However, in order to distinguish separable from entangled
states, we need a lower bound, which we will derive by generalizing
(\ref{conc_neu}), and which we will show to provide tight estimates of
mixed state concurrence. In particular, these estimates are {\em
  experimentally directly accessible}, through a small number of
projective measurements, for {\em arbitrary} $\varrho$.

Let us start from the observation that the two-fold copy
$\ket{\Psi}\otimes\ket{\Psi}$ is {\em symmetric} with respect to the
exchange of the copies of {\em both} subsystems.
This is the reason
why expectation values of $P_-\otimes P_+$ or $P_+\otimes P_-$, with
$P_+$ the symmetric counterpart of $P_-$, do not
contribute to pure state concurrence (\ref{conc_neu}),
and a finite expectation value of either observables unambiguously characterizes
a twofold copy of a state to be mixed.
A positive expectation values of $P_{-}\otimes P_{-}$ on the other hand
can be not only due to entanglement of the underlying state,
but also due to its mixing.

Combining these two observations, we conjecture that
\be
(c(\varrho))^2\ge\tr(\varrho\otimes\varrho\ V_{i})\ ,
\label{bound}
\ee
($i=1,2$) with
$V_{1}=4(P_{-}-P_{+})\otimes P_{-}$, and
$V_{2}=4P_{-}\otimes(P_{-}-P_{+})$.
The proof of this inequality is defered to the Appendix below. 

Inequality (\ref{bound}) implies the following important consequences:
\begin{itemize}
\item The lower bound on its right hand side can be expressed in terms of the
purities of $\varrho$, $\varrho_r^{(1)}$, and $\varrho_r^{(2)}$,
\begin{equation}
\tr(\varrho\otimes\varrho\ V_{i})
=2(\tr\varrho^2-\tr(\varrho_r^{(i)})^2)\ ,
\end{equation}
with $\varrho_r^{(i)}$ the reduced density matrix of either subsystem, in
close analogy to the expression
$c^2(\Psi)=2(1-\tr(\varrho_r^{(i)})^2)$
for pure state concurrence \cite{run01}.
\item There is an interesting interpretation of
$V$: Any non-vanishing
contribution to the expectation value on the right hand side of
(\ref{bound}), where the 
mixedness of the state is already substracted, must be due to
nonvanishing quantum correlations insribed in $\varrho$: Since the left hand side,
$c(\varrho)$, is strictly positive for entangled states, 
and vanishes exactly for separable states,
$\tr(\varrho\otimes\varrho V_{i})$ must be non-positive for any separable
density operator. Conversely, positive values of
$\tr(\varrho\otimes\varrho V_{i})$ 
unambiguously identify $\varrho$ as entangled. Consequently, $V$ qualifies as
a generalized entanglement witness, applicable for
{\em arbitrary} states,
without using a priori knowledge on $\varrho$.
\item Moreover, and most importantly, our new lower bound is given in terms of
  expectation values of $P_-$ and $P_+$. Hence it can be directly measured, as
the probabilities of finding the
two-fold copies of each 
state's individual subsystems with positive or negative parity. This implies 
only little overhead as compared to the experimental measurement of
pure state concurrence \cite{expmc}. 
\end{itemize}

\begin{figure*}[t]
\begin{center}
\includegraphics[width=0.95\textwidth]{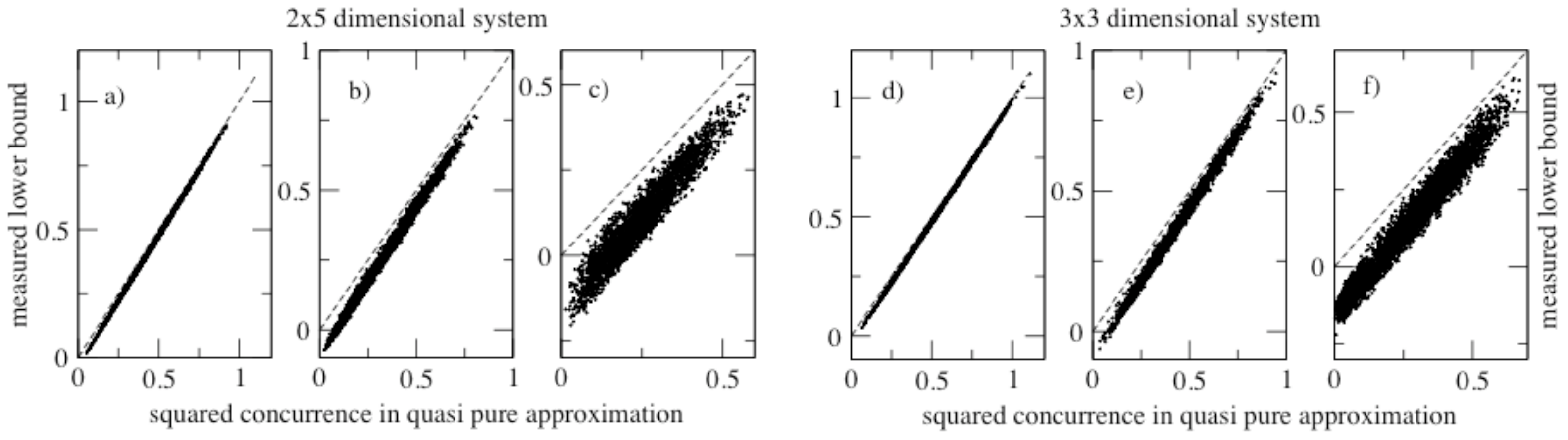}
\caption{
Squared mixed-state concurrence $c(\varrho)^2$, approximated
by its measurable lower bound
$\tr\left(\varrho\otimes\varrho\ (V_{1}+V_{2})/2\right)$ as given by (\ref{bound}), vs.
its lower bound in quasi pure approximation,
for random states
of a $2\times 5$ dimensional ($a$, $b$, and $c$),
and of a $3\times 3$ dimensional system ($d$, $e$, $f$).
The different panels represent states with different degrees of mixing:
$a$ and $d$ display the case of weakly mixed states
($0.2\le\sqrt{1-\tr\varrho^2}\le 0.21$),
$b$ and $e$ correspond to the regime of intermediate mixing
($0.4\le\sqrt{1-\tr\varrho^2}\le 0.405$),
and $c$ and $f$ show strongly mixed states
($0.529\le\sqrt{1-\tr\varrho^2}\le 0.533$).
The dashed lines indicate equality of both bounds.
In particular, for highly entangled states and for states with little
mixing the bound is very good;
but it also yields a surprisingly good characterization of rather
strongly mixed states.}
\label{fig}
\end{center}
\end{figure*}

Let us assess the tightness of (\ref{bound}), by  
evaluating the expectation value of $V$ 
for increasingly mixed random states of $2\times 5$ and
$3\times 3$ dimensional systems. The random states were obtained by 
random unitary evolution of a
pure state of a tripartite system, composed of the bipartite subspace
which supports the desired random states, and an environment
component, followed by a 
subsequent trace over the environment, at different times \cite{qp}.
The estimate (\ref{bound}) is then compared with the states'
concurrence in quasi pure approximation (qpa) \cite{qp}
in Fig.~\ref{fig}, for samples of $600000$ random states.
Qpa is known to provide 
very good 
approximations of a mixed state's
concurrence if the mixing is not too large (hence the label ``quasipure'').
Our new bound is seen to be only slightly weaker than qpa, as clearly
demonstrated by the present 
comparison.
Indeed, the comparison is in most cases excellent, in particular for
weakly mixed states. Only for a relatively small portion of relatively
strongly mixed (and weakly entangled) states does (\ref{bound}) take negative
values (while qpa 
remains positive), and thus provide an inconclusive result. 

Thus, Eq.~(\ref{bound}) provides a reliable and {\em directly measurable}
bound for an unknown state's concurrence. How do the
  necessary experimental resources scale with the system's size
  (determined by the individual dimensions of the factor
  spaces)? According to the explicit form of $V$, we have to determine
  the state's weight on the symmetric and antisymmetric subspaces of
  the two copies of each of its components. Consequently, for a
  bipartite state,
  with ${\rm id_i}=P^{(i)}_-+P^{(i)}_+$, $i=1,2$, two
measurements need to be
performed (here we assume that such
parity measurement can be performed by measuring one single
observable, independently of the subsystems' dimension).
This is in favourable
contrast to a tomographic measurement, where $d^4-1$ observables
need to be measured \cite{chuang00}, with $d$ the dimension of the
subsystems. 

Let us conclude with a brief digression on what we understand by a 
two-fold copy of an {\em unknown quantum state}, since this ansatz, while 
widely accepted \cite{phorodecki03,bovino:240407,pereswootters,brun04},
still raises some controversy in the literature \cite{quant-ph/0606017}: 
What we do assume here is the availability of a {\em reliable} source 
producing two faithful copies of the same state $\varrho$. Thus, the 
experimentalist who wants to implement our measurement scheme has to be 
certain about the source providing $\varrho\otimes\varrho$, but can be perfectly 
ignorant about the initialization of the source, and, hence, of the 
specific state $\varrho$ which is delivered in a two-fold copy. In itself, 
the preparation of two identical copies of $\varrho$ is a very realistic 
experimental task, given the stunning control over the microscopic 
constituents of matter, e.g., in state of the art quantum optical 
experiments.

We are indebted to
Luiz Davidovich,
Paulo Henrique Souto Ribeiro, and
Stephen Patrick Walborn
for fruitful discussions, comments and remarks.
This work was supported by a Feodor-Lynen fellowship
of Alexander von Humboldt foundation.

{\em Appendix:} For completeness, let us prove inequality (\ref{bound}):
we show that
the estimate applies for the average concurrence of {\em any} pure
state decomposition 
$\varrho =\sum_i\prj{\phi_i}$:
\bqa
\tr(\varrho\otimes\varrho V_{i})&=&
\sum_{ij}\bra{\phi_i}\otimes\bra{\phi_j}V_{i}\ket{\phi_i}\otimes\ket{\phi_j}\n\\
&\le&
\bigl(\sum_i\sqrt{\bra{\phi_i}\otimes\bra{\phi_i}A\ket{\phi_i}\otimes\ket{\phi_i}}\bigr)^2\n\\ 
&=&
\sum_{ij}c(\phi_i)c(\phi_j)\ ,
\label{glapp5}
\eqa
and thus, in particular, for the decomposition that achieves the
minimum average concurrence. 
It is sufficient to demonstrate the inequality's validity for each
single term in the above sum.
For convenience, we rebaptize $\ket{\phi_i}$ as
$\ket{\psi}$, and $\ket{\phi_j}$ as $\ket{\phi}$. Given $\ket{\phi}$'s
Schmidt decomposition 
$\ket{\phi}=\sum_i\sqrt{\lambda_i}\ket{i}\otimes\ket{i}$ and
$\ket{\psi}$'s expansion in the same one-particle bases,
$\ket{\psi}=\sum_{ij}\psi_{ij}\ket{i}\otimes\ket{j}$,
(\ref{glapp5}) can be written as
\bqa
\sum_{i\neq j}\psi_{ii}^\ast\psi_{jj}\sqrt{\lambda_i\lambda_j}-\left|\psi_{ij}\right|^2\lambda_i
&\le\n\\
\sqrt{\sum_{i\neq j}\lambda_i\lambda_j}
\sqrt{\sum_{{i\neq j}\atop{p\neq q}}
\left|\psi_{ip}\psi_{jq}-\psi_{iq}\psi_{jp}\right|^2}\ ,
\label{glapp10}
\eqa
for $V_{2}$, (and analogously for $V_{1}$).
With the help of the Cauchy-Schwarz inequality, the right hand side
(RHS) of this expression can be bounded from below,
\bqa
\mbox{RHS}&=&
\sqrt{\sum_{i\neq j}\lambda_i\lambda_j}
\sqrt{\sum_{{i\neq j}\atop{p\neq q}}
\left|\psi_{ip}\psi_{jq}-\psi_{iq}\psi_{jp}\right|^2}\n\\
&\ge&
\sqrt{\sum_{i\neq j}\lambda_i\lambda_j}
\sqrt{\sum_{i\neq j}
\left|\psi_{ii}\psi_{jj}-\psi_{ij}\psi_{ji}\right|^2}\n\\
&\ge&
\sum_{i\neq j}
\sqrt{\lambda_i\lambda_j}\left|\psi_{ii}\psi_{jj}-\psi_{ij}\psi_{ji}\right|\ .\n
\eqa
For the left hand side (LHS), we find
\bqa
\mbox{LHS}&=&\sum_{i\neq j}
\psi_{ii}^\ast\psi_{jj}\sqrt{\lambda_i\lambda_j}-\left|\psi_{ij}\right|^2\lambda_i\n\\
&=&\frac12\sum_{i\neq j}
\left(\psi_{ii}^\ast\psi_{jj}+\psi_{jj}^\ast\psi_{ii}\right)\sqrt{\lambda_i\lambda_j}-
\n\\&&\hspace{1cm}
\left(\left|\psi_{ij}\right|^2\lambda_{i}+\left|\psi_{ji}\right|^2\lambda_{j}\right)\n\\
&\le&\sum_{i\neq j}
\left(\left|\psi_{ii}\psi_{jj}\right|-\left|\psi_{ji}\psi_{ji}\right|\right)
\sqrt{\lambda_i\lambda_j}\n\\
&\le&
\sum_{i\neq j}\left|\psi_{ii}\psi_{jj}-\psi_{ij}\psi_{ji}\right|\sqrt{\lambda_i\lambda_j}\ ,\n
\eqa
where
$(|\psi_{ij}|\sqrt{\lambda_{i}}-|\psi_{ji}|\sqrt{\lambda_{j}})^{2}\ge 0$,
$\left|\psi_{ij}\right|^2+\left|\psi_{ji}\right|^2\ge
2\left|\psi_{ij}\psi_{ji}\right|$,
$\psi_{ii}^\ast\psi_{jj}+\psi_{jj}^\ast\psi_{ii}=2\Re(\psi_{ii}^\ast\psi_{jj})\le
2\left|\psi_{ii}\psi_{jj}\right|$,
and the triangle inequality
$\left|\psi_{ii}\psi_{jj}\right|-\left|\psi_{ij}\psi_{ji}\right|\le
\left|\psi_{ii}\psi_{jj}-\psi_{ij}\psi_{ji}\right|$ were used. Thus,
${\rm LHS}\leq{\rm RHS}$, what was our initial claim.

\bibliography{../../../referenzen}

\end{document}